\documentclass[prl,twocolumn,showpacs]{revtex4}
\usepackage{amsmath, amsgen,amstext,amsbsy,amsopn,amsthm, amssymb}
\theoremstyle{definition}

\newcommand{\eps}{\epsilon}

\newcommand{\aoo}{{a_{\bf 0}^{\phantom *}}}
\newcommand{\aos}{a_{\bf 0}^*}
\newcommand{\ak}{{a_{\bf k}^{\phantom *}}}
\newcommand{\aks}{a_{\bf k}^*}
\newcommand{\aq}{a_{\bf q}}

\newcommand{\no}{n_{\bf 0}}
\newcommand{\kk}{{\bf k}}
\newcommand{\pp}{{\bf p}}
\newcommand{\qq}{{\bf q}}
\newcommand{\hi}{\mathcal{H}}
\newcommand{\hip}{{\mathcal{H}'}}

\newcommand{\half}{\mbox{$\frac{1}{2}$}}

\newcommand{\x}{{\bf r}}

\newcommand{\Tr}{{\rm Tr} }

\begin{document}
\title{Justification of $c$-Number Substitutions in Bosonic Hamiltonians}
\author{Elliott H.~Lieb}
\affiliation{Department of Physics,
Jadwin Hall, Princeton University,
P.~O.~Box 708, Princeton, New Jersey 08544}
\author{Robert Seiringer}
\affiliation{Department of Physics,
Jadwin Hall, Princeton University,
P.~O.~Box 708, Princeton, New Jersey 08544}
\author{Jakob Yngvason}
\affiliation{Institut f\"ur Theoretische Physik,
Universit\"at
Wien, Boltzmanngasse 5, A-1090 Vienna, Austria}

\date{Feb. 28, 2005}

\begin{abstract}
  The validity of substituting a $c$-number $z$ for the $\kk={\bf 0}$
  mode operator $\aoo$ is established rigorously in full generality,
  thereby verifying one aspect of Bogoliubov's 1947 theory. This
  substitution not only yields the correct value of thermodynamic
  quantities like the pressure or ground state energy, but also the
  value of $|z|^2$ that maximizes the partition function equals the
  true amount of condensation in the presence of a gauge-symmetry
  breaking term --- a point that had previously been elusive.
\end{abstract}


\maketitle

One of the key developments in the theory of the Bose gas, especially
the theory of the low density gases currently at the forefront of
experiment, is Bogoliubov's 1947 analysis \cite{bog} of the many-body
Hamiltonian by means of a $c$-number substitution for the most relevant
operators in the problem, the zero-momentum mode operators,
namely $\aoo\to z, \, \aos\to z^*$.  Naturally, the appropriate value
of $z$ has to be determined by some sort of consistency or variational
principle, which might be complicated, but the concern, expressed by
many authors over the years, is whether this sort of substitution is
legitimate, i.e., error free.  We address this latter problem here and
show, by a simple but rigorous analysis, that it is so under very
general circumstances.

The rigorous justification for this substitution, as far as calculating
the pressure is concerned, was done in a classic paper of Ginibre
\cite{gin} in 1968, but it does not seem to have percolated into the
general theory community. In textbooks it is often said, for instance,
that
it is tied to the imputed `fact' that the expectation value of the number
operator $\no =\aos \aoo$ is of order $V=$ volume. (This was the argument
in \cite{bog}). That is, Bose-Einstein condensation (BEC)  justifies the
substitution. As Ginibre pointed out, however, BEC has nothing to do with
it. The $z$ substitution still gives the right answer even if $\no$ is
small (but it is a useful calculational tool only if $\no$ is
macroscopic). Thus, despite \cite{gin} and the thorough review of these
matters
in \cite{zag}, there is some confusion in the literature and clarification
could
be useful.

In this short note we do three things. 1.) We show how Ginibre's
result can be easily obtained in a few simple lines. While he used
coherent states, he did not use the Berezin-Lieb inequality
\cite{BL,Li,simon}, derived later, which efficiently gives upper bounds.
This inequality gives explicit error bounds which, typically, are only
order one compared to the total free energy or pressure times volume,
which are order $N=$ particle number. \

2.)  This allows us to go beyond \cite{gin} and make $c$-number
substitutions for {\it many} $\kk$-modes at once, provided the number
of modes is lower order than $N$.  \

3.)  We show how the optimum value of $z$ yields, in fact, the
expectation value $\langle\no\rangle$ in the true state when a gauge
breaking term is added to the Hamiltonian. More precisely, in the
thermodynamic limit (TL) the $|z|^2$ that maximizes the partition
function equals $|\langle \aoo \rangle | ^2$ and this equals $\langle
\no \rangle$, which is the amount of condensation --- a point that was
not addressed in full generality in previous work
\cite{gin,zag,BSP}.  The second of these equalities has previously
only been treated under some additional assumptions \cite{fannes} or
for some simplified models \cite{zag,suto}.  \footnote
{{\it Note added in proof:} After we submitted this paper A.\ S\"ut\H o
presented
a different proof of item 3 (math-ph/0412056).}

While we work here at positive temperature
$k_{\rm B}T=1/\beta$, our methods also work for the ground state (and
are even simpler in that case).
To keep this note short and, hopefully, readable, we will be a bit
sketchy in places but there is no difficulty filling in the details.

The use of coherent states \cite{klauder,feng} to give accurate upper
and lower bounds to energies, and thence to expectation values, is
effective in a wide variety of problems \cite{liebcoh}, e.g., quantum
spin systems in the large $S$ limit \cite{Li}, the Dicke model
\cite{HL}, the strong coupling polaron \cite{LT}, and the proof that
Thomas-Fermi theory is exact in the large atom limit \cite{li2,thir}.
For concreteness and relevance, we concentrate on the Bose gas problem
here, and we discuss only the total, correct Hamiltonian.
Nevertheless, the same conclusions hold also for variants, such as
Bogoliubov's truncated Hamiltonian (the ``weakly imperfect Bose gas''
\cite{zag,bog}) or other modifications, provided we are in the
stability regime (i.e., the regime in which the models make sense).
We are not
claiming that any particular approximation is valid. That is a completely
different story that has to be decided independently.
The method can also be modified to incorporate inhomogeneous systems.
The message is the same in all cases, namely that the $z$ substitution
causes no errors (in the TL), even if there is no BEC, whenever it is
applied to
physically stable systems.  Conversely, if the system is stable after
the $z$ substitution then so is the original one.

We start with the well-known Hamiltonian for bosons in a large box of
volume $V$, expressed in terms of the second-quantized creation and
annihilation operators $\ak, \aks$ satisfying the canonical
commutation relations,
\begin{equation}
H = \sum_\kk k^2 \aks \ak + \frac{1}{2V}\sum_{\kk, \pp, \qq}
\nu(\pp) a_{\kk + \pp}^* a_{\qq - \pp}^* \ak \aq  ,
\end{equation}
(with $\hbar = 2m =1$). Here, $\nu$ is the Fourier transform of the
two-body potential $v(\x)$.  We assume that there is a bound on the
Fourier coefficients $|\nu(\kk)| \leq \varphi <\infty$.

The case of
hard core potentials can be taken care of in the following way. First cut
off the hard core potential $v$ at a height $10^{12}$ eV. It is easy
to prove, by standard methods, that this cutoff will have a
negligible effect on the exact answer.
After the cutoff $\varphi$
will be about $10^{12}$ eV \r{A}$^3$, and according to what we prove
below, this substitution will affect the chemical potential only by
about $\varphi /V$, which is truly negligible when $V= 10^{23}$
\r{A}$^3$.

If we replace the operator $\aoo$ by a complex number $z$ and $\aos$
by $z^*$ everywhere in $H$ we obtain a Hamiltonian $H'(z)$ that acts
on the Fock-space of all the modes other than the $\aoo$ mode.
Unfortunately, $H'(z)$ does not commute with the particle number
$N^>\equiv\sum_{\kk\neq {\bf 0}} \aks \ak$. It is convenient, therefore,
to work in
the grand canonical ensemble and consider $H_\mu = H- \mu N = H-\mu (
\aos \aoo + N^>)$ and, correspondingly, $H^\prime_\mu(z)= H'(z) -\mu
(|z|^2 +
N^>)$.

The partition functions are  given by
\begin{align}\label{partition}
e^{\beta V p(\mu)} \equiv \Xi(\mu) &= \Tr_\hi \exp[ -\beta H_\mu] \\
e^{\beta V p'(\mu)} \equiv\Xi'(\mu) &= \int d^2\!z\, \Tr_\hip
\exp[ -\beta H_\mu^\prime(z)]
\label{partitionint}
\end{align}
where $\hi$ is the full Hilbert (Fock) space, $\hip$ is the Fock space
without
the $\aoo$ mode, and $d^2\!z\,\equiv \pi^{-1} dxdy$ with $z=x+iy$. The
functions  $p(\mu)$
and $p'(\mu)$ are the corresponding finite volume pressures.

The pressure $p(\mu)$ has a finite TL for all $\mu<
\mu_{\rm critical}$, and it is a convex function of $\mu$. For the
non-interacting gas, $\mu_{\rm critical}=0$, but for any {\it realistic}
system
$\mu_{\rm critical}=+\infty$. In any case, we assume $\mu< \mu_{\rm
  critical}$, in which case both the pressure and the density are finite.

Let $|z\rangle = \exp\{-|z|^2/2 +z \aos \}\, |0\rangle$ be the
coherent state vector in the $\aoo$ Fock space and let $\Pi(z)
=|z\rangle\langle z|$ be the projector onto this vector. There are six
relevant operators containing $\aoo$ in $H_\mu$, which have the following
expectation
values \cite{klauder} (called {\it lower symbols})
\begin{align}\nonumber
\langle z| \aoo  |z \rangle &= z,
& \!\!\!\langle z| \aoo \aoo |z  \rangle &= z^2, &\!\!\!
\langle z| \aos \aoo |z  \rangle &= |z|^2 \\
\langle z| \aos  |z \rangle &= z^*, & \!\!\!
 \langle z| \aos \aos  |z  \rangle & = z^{*2}, & \!\!\!
\langle z|  \aos \aos \aoo \aoo |z  \rangle &= |z|^4. \nonumber
\end{align}
Each also has an {\it upper symbol}, which is a function of $z$ (call
it $u(z)$ generically) such that an operator $F$ is represented as $F
= \int d^2\!z\, u(z) \Pi(z)$. These symbols are
\begin{align}
\aoo &\to z, & \aoo\aoo &\to z^2, &  \aos \aoo &\to |z|^2-1 \nonumber \\
 \aos &\to z^*, & \!\!\!  \aos \aos &\to z^{*2}, &\!\!\!  \aos \aos \aoo
\aoo
 &\to |z|^4 -4|z|^2 +2. \nonumber
\end{align}

It will be noted that the operator $H^\prime_\mu(z)$, defined above,
is obtained from $H_\mu$ by substituting the lower symbols for the six
operators.  If we substitute the upper symbols instead into $H_\mu$ we
obtain a slightly different operator, which we write as $H^{\prime
  \prime}_\mu(z) = H^\prime_\mu(z) + \delta_\mu(z)$ with
\begin{multline}\label{delta}
 \delta_\mu(z) = \mu+ \frac{1}{2V}\Big[ (-4|z|^2 +2) \nu({\bf 0}) \\
- \sum_{\kk \neq {\bf 0}}
a_{\kk}^*a_{\kk}^{\phantom *}\big(2 \nu({\bf 0}) + \nu(\kk)+
\nu(-\kk)\big) \Big] .
\end{multline}

The next step is to mention two inequalities, of which the first is
\begin{equation}\label{lowerbound}
\Xi(\mu) \geq \Xi'(\mu)\,.
\end{equation}
This is a consequence of the following two facts: The completeness
property of coherent states,
$ \int d^2\!z\, \Pi (z) = {\rm Identity}$, and
\begin{equation} \label{jensen}
\langle z\otimes \phi | e^{-\beta H_\mu} | z\otimes \phi \rangle \geq
e^{-\beta \langle z\otimes \phi | H_\mu |z \otimes \phi \rangle } =
 e^ {-\beta \langle \phi |H_\mu^\prime ( z) |\phi \rangle },
\end{equation}
where $\phi$ is any normalized vector in $\hip$. This
 is Jensen's inequality for the expectation value of a
convex function (like the exponential function) of an operator.

To prove (\ref{lowerbound}) we take $\phi$ in (\ref{jensen}) to be one
of the normalized eigenvectors of $H_\mu^\prime (z)$, in which case
$\exp\{\langle \phi |-\beta H_\mu^\prime ( z) |\phi \rangle\} =
\langle \phi | \exp\{-\beta H_\mu^\prime ( z) \}|\phi \rangle $.  We
then sum over all such eigenvectors (for a fixed $z$) and integrate
over $z$.  The left side is then $\Xi(\mu)$, while the right side is
$\Xi'(\mu)$.

The second inequality \cite{BL,Li,simon} is
\begin{equation}\label{upper}
\Xi(\mu) \leq \Xi''(\mu)\equiv e^{\beta V p''(\mu)},
\end{equation}
where $ \Xi''(\mu)$ is similar to $ \Xi'(\mu)$ except that
$H_\mu^\prime(z)$ is replaced by $H_\mu^{\prime \prime}(z)$.  Its
proof is the following. Let $|\Phi_j \rangle \in \hi$ denote the
complete set of normalized eigenfunctions of $H_\mu$. The partial
inner product $|\Psi_j(z)\rangle = \langle z| \Phi_j\rangle $ is a
vector in $\hip$ whose square norm, $c_j(z) = \langle \Psi_j (z) |
\Psi_j(z) \rangle_\hip$ satisfies $\int d^2\!z\, c_j(z) =1$. By using the
upper symbols, we can write $\langle \Phi_j | H_\mu | \Phi_j\rangle =\int
d^2\!z\, \langle \Psi_j (z) | H_\mu^{\prime \prime} (z) | \Psi_j
(z)\rangle = \int
d^2\!z\,\langle \Psi_j' (z) | H_\mu^{\prime \prime} (z)|
\Psi_j'(z)\rangle c_j(z) $, where $|\Psi_j'(z) \rangle$ is the
normalized vector $c_j(z)^{-1/2} \Psi_j(z)$.  To compute the trace, we
can exponentiate this to write $\Xi(\mu)$ as
\begin{equation}\nonumber
 \sum_j \exp\left\{-\beta \int d^2\!z\, c_j(z)
\langle \Psi_j' (z) | H_\mu^{\prime \prime} (z)|
\Psi_j'(z)\rangle  \right\}.
\end{equation}
Using Jensen's inequality twice, once for functions and once for
expectations as in (\ref{jensen}), $\Xi(\mu)$ is less than
\begin{multline} \nonumber
 \sum_j \int d^2\!z\, c_j(z) \exp\left\{
\langle \Psi_j' (z) | -\beta H_\mu^{\prime \prime} (z) |
\Psi_j'(z)\rangle \right\} \\ \leq \sum_j \int d^2\!z\, c_j(z)
\langle \Psi_j' (z) | \exp\left\{-\beta H_\mu^{\prime \prime} (z) \right\}
|
\Psi_j'(z)\rangle .
\end{multline}
Since $\Tr\, \Pi(z) = 1$, the last
expression can be rewritten
\begin{equation}\nonumber
\int d^2\!z\, \sum_j \langle \Phi_j |  \Pi(z) \otimes
\exp\left\{-\beta H_\mu^{\prime \prime}(z)
\right\}| \Phi_j \rangle = \Xi''(\mu).
\end{equation}

Thus, we have that
\begin{equation}\label{correx0}
\Xi'(\mu) \leq \Xi(\mu) \leq \Xi''(\mu).
\end{equation}
The next step is to try to relate $ \Xi''(\mu)$ to $\Xi'(\mu) $. To
this end we have to bound $\delta_\mu(z)$ in (\ref{delta}). This is
easily done in terms of the total number operator whose lower symbol
is $N^{\prime}(z) = |z|^2 + \sum_{\kk \neq {\bf 0}}\aks \ak$. In
terms of the bound $\varphi$ on $\nu(\pp)$
\begin{equation}\label{bounddelta}
|\delta_\mu(z)|  \leq 2\varphi (N'(z)+\half )/V +|\mu| \ .
\end{equation}
Consequently, $ \Xi''(\mu)$ and
$\Xi'(\mu) $ are related by the inequality
\begin{equation}
\Xi''(\mu) \leq \Xi'(\mu + 2\varphi/V) e^{\beta (|\mu|+\varphi/V)}.
\label{correx1}
\end{equation}
Equality of the pressures $p(\mu)$, $p'(\mu)$ and $p''(\mu)$ in the TL
follows from (\ref{correx0}) and (\ref{correx1}).

Closely related to this point is the question of relating $\Xi(\mu)$
to the maximum value of the integrand in (\ref{partitionint}), which
is $\max_z \Tr_\hip \exp[ -\beta H_\mu^\prime(z)] \equiv e^{\beta V
  p^{\max}}$.  This latter quantity is often used in discussions of
the $z$ substitution problem, e.g., in refs. \cite{gin,zag}.  One
direction is not hard.  It is the inequality (used in ref. \cite{gin})
\begin{equation}\label{junk}
\Xi(\mu) \geq \max_z \Tr_\hip  \exp[ -\beta H_\mu^\prime(z)],
\end{equation}
and the proof is the same as the proof of (\ref{lowerbound}), except
that this time we replace the completeness relation for the coherent
states by the simple inequality ${\rm Identity} \geq \Pi(z)$ for
any fixed number $z$.

For the other direction, split
the integral defining $\Xi''(\mu)$ into a part where $|z|^2 <  \xi$ and
$|z|^2\geq \xi$. Thus,
\begin{multline}\label{cheb}
  \Xi''(\mu) \leq  \xi \max_z \Tr_\hip  \exp[ -\beta
H_\mu^{\prime\prime}(z)]
\\ +  \frac 1{\xi} \int_{|z|^2\geq \xi} d^2\!z\,
  |z|^2 \, \Tr_\hip \exp[ -\beta H_\mu^{\prime\prime}(z)].
\end{multline}
Dropping the condition $|z|^2\geq \xi$ in the last integral and using
$|z|^2\leq N'(z)=N''(z)+1$, we see that the second line in (\ref{cheb}) is
bounded above by $\xi^{-1} \Xi''(\mu) [V\rho''(\mu)+1]$, where
$\rho''(\mu)$ denotes the density in the $H_\mu''$ ensemble. Optimizing
over $\xi$  leads to
\begin{equation}\label{morejunk}
\Xi''(\mu) \leq 2 [V \rho''(\mu)+1]  \, \max_{z} \Tr_\hip  \exp[ -\beta
H_\mu^{\prime\prime}(z)].
\end{equation}
Note that $\rho''(\mu)$ is order one, since $p''(\mu)$ and $p(\mu)$
agree in the TL (and are convex in $\mu$), and we assumed that the density
in the original
ensemble is finite. By (\ref{bounddelta}), $H_\mu^{\prime\prime}\geq
H_{\mu+2\varphi/V}^\prime-|\mu|-\varphi/V$, and  it follows from
(\ref{upper}), (\ref{morejunk}) and~(\ref{junk}) that $p^{\max}$
agrees with the true pressure $p$ in the TL. Their
difference, in fact, is at most $O(\ln V /V)$. This is the result obtained
by
Ginibre in \cite{gin} by more complicated arguments, under the
assumption of superstability of the interaction, and without the
explicit error estimates obtained here.

To summarize the situation so far, we have four expressions for the
grand canonical pressure. They are all equal in the TL
limit,
\begin{equation}\label{equality}
p(\mu) = p'( \mu) = p''(\mu) =  p^{\max}(\mu)
\end{equation}
when $\mu$ is not a point at which the density can be infinite.

Our second main point is that not only is the $z$ substitution valid
for $\aoo$ but it can also be done for many modes simultaneously.  As
long as the number of modes treated in this way is much smaller than
$N$ the effect on the pressure will be negligible.  Each such
substitution will result in an error in the chemical potential that is
order $\varphi /V$. The proof of this fact just imitates what was done
above for one mode.  Translation invariance is not important here; one
can replace any mode such as $\sum_{\bf k} g^{\phantom *}_{\bf k} \ak$
by a $c$-number, which can be useful for inhomogeneous systems.

A more delicate point is our third one, and it requires, first, a
discussion of the meaning of `condensate fraction' that goes beyond
what is usually mentioned in textbooks, but which was brought out in
\cite{bog,griffiths,roepstorff}. The `natural' idea would be to
consider $V^{-1}\langle \no \rangle$. This, however, need not be a
reliable measure of the condensate fraction for the following reason.
If we expand $\exp \{-\beta H\}$ in eigenfunctions of the number
operator $\no$ we would have $\langle \no \rangle = \sum_n n
\gamma (n)$, where $\gamma (n)$ is the probability that $\no=n$. One
would like to think that $\gamma(n) $ is sharply peaked at some
maximum $n$ value, but we do not know if this is the case. $\gamma(n)$
could be flat, up to the maximum value or, worse, it could have a
maximum at $n=0$.
Recall that precisely this happens for the
Heisenberg quantum ferromagnet \cite{griffiths}; by virtue of
conservation of total spin angular momentum, the distribution of
values of the $z$-component of the total spin, $S^z$, is a strictly
decreasing function of $|S^z|$. Even if it were flat, the expected
value of $S^z$ would be half of the spontaneous magnetization that one
gets by applying a weak magnetic field.

With this example in mind, we see that the only physically reliable
quantity is $\lim_{\lambda \to 0} \lim_{V \to \infty} V^{-1}\langle
\no \rangle_{\mu, \lambda}$, where the expectation is now with respect
to a Hamiltonian $H_{\mu,\lambda}= H_\mu + \sqrt{V}(\lambda \aoo
+\lambda^*\aos)$ \cite{bog}. Without loss of generality, we assume
$\lambda$ to be real.  We will show that for almost every $\lambda$,
the density $\gamma(V\rho_0)$ converges in the TL to a $\delta$-function
at
the point $\widehat\rho_0=\lim_{V\to\infty} |z_{\rm max}|^2/V$, where
$z_{\rm
  max}$ maximizes the partition function $\Tr_{\hip} \exp\{ -\beta
H'_{\mu,\lambda}(z)\}$.  That is,
\begin{equation}\label{19}
 V^{-1} \langle \no\rangle_{\mu,\lambda}=
 V^{-1} |\langle \aoo\rangle_{\mu,\lambda}|^2
=V^{-1} |z_{\max} |^2
\end{equation}
in the TL. This holds for those $\lambda$ where the pressure in the TL
is differentiable; since $p(\mu,\lambda)$ is convex (upwards) in
$\lambda$ this is true almost everywhere. The right and left
derivatives exist for every $\lambda$ and hence
$\lim_{\lambda \to 0+} \lim_{V\to\infty} V^{-1} |\langle
\aoo\rangle_{\mu,\lambda}|^2$ exists.

The expectation values $\langle \no\rangle_{\mu,\lambda}$ and $\langle
\aoo\rangle_{\mu,\lambda}$ are obtained by integrating $(|z|^2-1)$ and
$z$, respectively, with the weight $W_{\mu,\lambda}(z)\equiv
\Xi(\mu,\lambda)^{-1} \Tr_{\hip} \langle z| \exp\{-\beta
H_{\mu,\lambda}\}| z\rangle$. We will show that this weight converges
to a $\delta$-function at $z_{\max}$ in the TL, implying (\ref{19}).
If we could replace $W_{\mu,\lambda}(z)$ by
$W_{\mu,0}(z)e^{-\beta \lambda\sqrt{V}(z+z^*)}$, this would follow from
Griffiths'
argument \cite{griffiths} (see also \cite[Sect.~1]{DLS}). Because
$[H,\aoo]\neq 0$, $W_{\mu,\lambda}$ is not of this product form.
However, the weight for $\Xi''(\mu,\lambda)$, which is
$W''_{\mu,\lambda}(z)\equiv \Xi''(\mu,\lambda)^{-1}
\Tr_{\hip}\exp\{-\beta H^{\prime\prime}_{\mu,\lambda}(z)\}$, does have the
right form. In the following we shall show that the two weights are
equal apart from negligible errors.

Equality (\ref{equality}) holds also for all $\lambda$, i.e.,
$p(\mu,\lambda) = p''( \mu,\lambda) = p^{\max}(\mu,\lambda)$ in the
TL.  In fact, since the upper and lower symbols agree for $\aoo$ and
$\aos$, the error estimates above remain unchanged.  (Note that since
$\sqrt{V}|\aoo+\aos|\leq \delta (N+\half) + V/\delta$ for any $\delta>0$,
$p(\mu,\lambda)$ is finite for all $\lambda$ if it is finite for
$\lambda=0$ in a small interval around $\mu$.) At any point of
differentiability with respect to $\lambda$, Griffiths' theorem
\cite{griffiths} (see \cite[Cor.~1.1]{DLS}), applied to the partition
function $\Xi''(\lambda,\mu)$, implies that $W''_{\mu,\lambda}(\zeta \sqrt
V)$ converges to a $\delta$-function at some point $\widehat \zeta$ on the
real axis as $V\to\infty$. (The original Griffiths argument can easily
be extended to two variables, as we have here. Because of radial
symmetry, the derivative of the pressure with respect to ${\rm Im\,
}\lambda$ is zero at any non-zero real $\lambda$.)  Moreover, by
comparing the derivatives of $p''$ and $p^{\max}$ we see that
$\widehat \zeta= \lim_{V\to\infty} z_{\rm max}/\sqrt{V}$, since $z_{\rm
  max}/\sqrt{V}$ is contained in the interval between the left and
right derivatives of $p^{\max}(\mu,\lambda)$ with respect to
$\lambda$.

We shall now show that the same is true for $W_{\mu,\lambda}$. To this
end, we add another term to the Hamiltonian, namely $\eps F\equiv \eps V
\int
d^2\!z\, \Pi(z) f(zV^{-1/2})$, with $\eps$ and $f$ real. If $f(\zeta)$ is
a
nice function of two real variables with bounded second
derivatives, it is then easy to see that the upper and lower symbols of
$F$ differ only by a term of order $1$. Namely, for some $C>0$
independent of $z_0$ and $V$,
\begin{equation}\nonumber
\left| V \int d^2\!z\, |\langle z|z_0\rangle |^2
\left( f(zV^{-1/2})- f(z_0 V^{-1/2})\right) \right| \leq C .
\end{equation}
Hence, in particular,
$p(\mu,\lambda,\eps)=p''(\mu,\lambda,\eps)$ in the TL. Moreover, if
$f(\zeta)=0$ for $|\zeta-\widehat \zeta|\leq \delta$, then the pressure is
independent of $\eps$ for $|\eps|$ small enough (depending only on
$\delta$). This can be seen as follows. We have
\begin{equation}\label{21}
p''(\mu,\lambda,\eps) - p''(\mu,\lambda,0) = \frac 1{\beta V} \ln
\left\langle
e^{-\beta\eps V f(zV^{-1/2})} \right\rangle,
\end{equation}
where the last expectation is in the $H_\mu''$ ensemble at $\eps =0$.  The
corresponding distribution is exponentially localized at
$z/\sqrt{V}=\widehat
\zeta$ \cite{griffiths,DLS}, and therefore the right side of (\ref{21})
goes to zero in the TL for small enough $\eps$.  In particular, the
$\eps$-derivative of the TL pressure at $\eps=0$ is zero.  By
convexity in $\eps$, this implies that the derivative of $p$ at finite
volume, given by $V^{-1}\langle F\rangle_{\mu,\lambda} = \int
d^2\!z\,f(zV^{-1/2}) W_{\mu,\lambda}(z)$, goes to zero in the TL.
Since $f$ was arbitrary, $V\int_{|\zeta-\widehat \zeta|\geq \delta}
d^2\!\zeta\,
W_{\mu,\lambda}(\zeta\sqrt V)\to 0$ as $V\to \infty$. This holds for all
$\delta>0$, and therefore proves the statement.

Our method also applies to the case when the pressure is not
differentiable in $\lambda$ (which is the case at $\lambda=0$ in the
presence of BEC). In this case, the resulting weights
$W_{\mu,\lambda}$ and $W''_{\mu,\lambda}$ need not be
$\delta$-functions, but Griffiths' method \cite{griffiths,DLS} implies
that they are, for $\lambda\neq 0$, supported on the real axis between
the right and left derivative of $p$ and, for $\lambda=0$, on a disc
(due to the gauge symmetry) with radius determined by the right
derivative at $\lambda=0$.  This, together with convexity, implies
that $\langle \no\rangle_\lambda$ is monotone increasing in $\lambda$
in the TL and, in particular, $\lim_{V\to\infty} V^{-1} \langle
\no\rangle_{\mu,\lambda=0}\leq \lim_{\lambda\to 0} \lim_{V\to \infty}
V^{-1} |\langle \aoo\rangle_{\mu,\lambda}|^2$, a fact which is
intuitively clear but has, to the best of our knowledge, not been
proved so far \cite{roepstorff} in this generality.  In fact, the only
hypothesis entering our analysis, apart from the bound $\varphi$ on
the potential, is the existence of the TL of the pressure and the
density.

We thank V.A. Zagrebnov for helpful comments. 
The work was supported in part by US NSF grants PHY 0139984-A01 (EHL)
and PHY 0353181 (RS), by an A.P.~Sloan Fellowship (RS), by EU grant
HPRN-CT-2002-00277 (JY) and by FWF grant P17176-N02 (JY).

\end{document}